# Time dependences of atmospheric Carbon dioxide fluxes

Riccardo DeSalvo [1,2]
[1] *California State University, 18111 Nordhoff Street, Northridge, CA 91330-8332 (USA).*
[2] *University of Sannio, Corso Garibaldi 107, Benevento 82100 (Italy).*
*e-mail address: riccardo.desalvo@gmail.com*

A mass conservation analysis generates tight estimations for the atmosphere's $CO_2$ retention time constant. It uses a simple leaky integrator model, abstracting from the many sources and sinks exchanging with the atmosphere. The well documented $CO_2$ deficit (the atmosphere retains only a fraction of the $CO_2$ injected by combustion of fossil fuels) combined with the exponential growth of fossil fuel use reveals a maximum characteristic time of less than 23 year for the transfer of $CO_2$ to a segregation sink. A lower limit of 18 year is provided by the rapid disappearance of $^{14}C$ after the ban of atmospheric atomic bomb tests. The oscillation amplitudes of atmospheric $CO_2$ concentration are also analyzed, showing correlation with landmass distribution and giving indications on the stability of long-term sinks. It is predicted that if the increase of anthropogenic production is halted, the $CO_2$ concentration will stabilize within a quarter century.

## I. INTRODUCTION

Understanding the cycle of $CO_2$ in the atmosphere is critical for predictions regarding future climate changes. The largest exchanges of atmospheric $CO_2$ are those with the biota, responsible for the seasonal oscillations with injection and sink rates at least an order of magnitude larger than all human contributions [1, pg 471-473 and references therein]. Anthropogenic production of $CO_2$ has been growing exponentially in the last two and a half centuries. It is still a small fraction of the natural cycles [2], but its accumulation is relevant. The industrial injection rates have been tabulated since 1750 [3], and are known with higher precision than the much larger natural sources and sinks. It has been known for long time that less than half of the $CO_2$ injected by industrial activities is actually found in the atmosphere, while the rest is sunk elsewhere. Evaluations of the actual human contribution to atmospheric $CO_2$ by adding up all sources and sinks are subject to large errors, due to the fact that they are the result of the subtraction of large numbers, each with its own uncertainty. The uncertainties are large enough that, unlike for all other gases, [1 table 2.1 page 166] lists no lifetime for atmospheric $CO_2$. It is shown here that it is not necessary to know in detail all the flows and sink mechanisms to precisely evaluate the retention time scale of $CO_2$ in the atmosphere. The atmosphere acts as a natural integrator, and the $CO_2$ concentration, precisely measured since 1957 [1], is a direct measurement of the net flow into the atmosphere, and of its retention capacity. The following mathematical analysis, based on global data, produces tight and independent upper and lower limits for the atmospheric $CO_2$ lifetime.

## II. MODEL

The model hinges on three key observations:

1. The exponential growth of atmospheric $CO_2$ tracks in exponential shape, but not in amplitude, the anthropogenic emissions.
2. A comparison of the tabulated production of $CO_2$ from combustion of fossil fuels with those of actual atmospheric concentration measurements, shows that less than 40% of $CO_2$ injected is retained while more than 60% is continuously shed.
3. Large seasonal oscillations of atmospheric $CO_2$ concentration have amplitude proportional to the total $CO_2$ inventory, and not to the amount of anthropogenic injection, while amplitude and phase depend on latitude.

The observed behavior can be modeled as a leaky integrator driven by anthropogenic injections and by a natural seasonal oscillator. A straightforward analysis of this model, combining point 1 and 2, reveals a relatively short characteristic time of less than 23 year for the transfer of $CO_2$ from the atmosphere to a long term segregation sink. The rapid disappearance of $^{14}C$ after the ban of atmospheric atomic bomb tests provides a lower limit of 18 year for this time constant. Study of point 3 indicates a strong, fast, time-varying flux in and out of the atmosphere, with an exchange time constant of little more than one month. The correlation of oscillation amplitude with landmass distribution confirms that this oscillation likely has a predominantly terrestrial origin.

## III. DATA DISCUSSION

The analysis uses public data provided by CDIAC [3]. The amount of $CO_2$ produced by industrial carbon consumption has been normalized to the mass of the atmosphere to get annual injection values in ppm. A baseline value of 283 ppm was added to account for the pre-industrial-revolution concentration levels, to allow for direct comparison with the measured atmospheric $CO_2$ concentration data. In spite of the vagaries of history, both curves are fit well by an exponential growth. The production of $CO_2$ in the last 250 years closely and consistently follows an exponential curve with a time constant $\tau_{inj}$ = 34.5±0.2 year (34.9±0.4 year if the fit is limited to the last 50 years). The measured $CO_2$ concentration follows a similar law with a time scale $\tau_{growth}$ = 43.3±0.4 year (the fit excludes the lower-quality, pre-1957 ice-core measurements, which may be biased or just too noisy). The deforestation contribution is also growing exponentially with a 170 to 190 year time constant; it provides a reasonable explanation for the difference between $\tau_{inj}$ and $\tau_{growth}$, which should otherwise be identical in a linear system. The deforestation data set has large systematic errors, which make its use problematic.

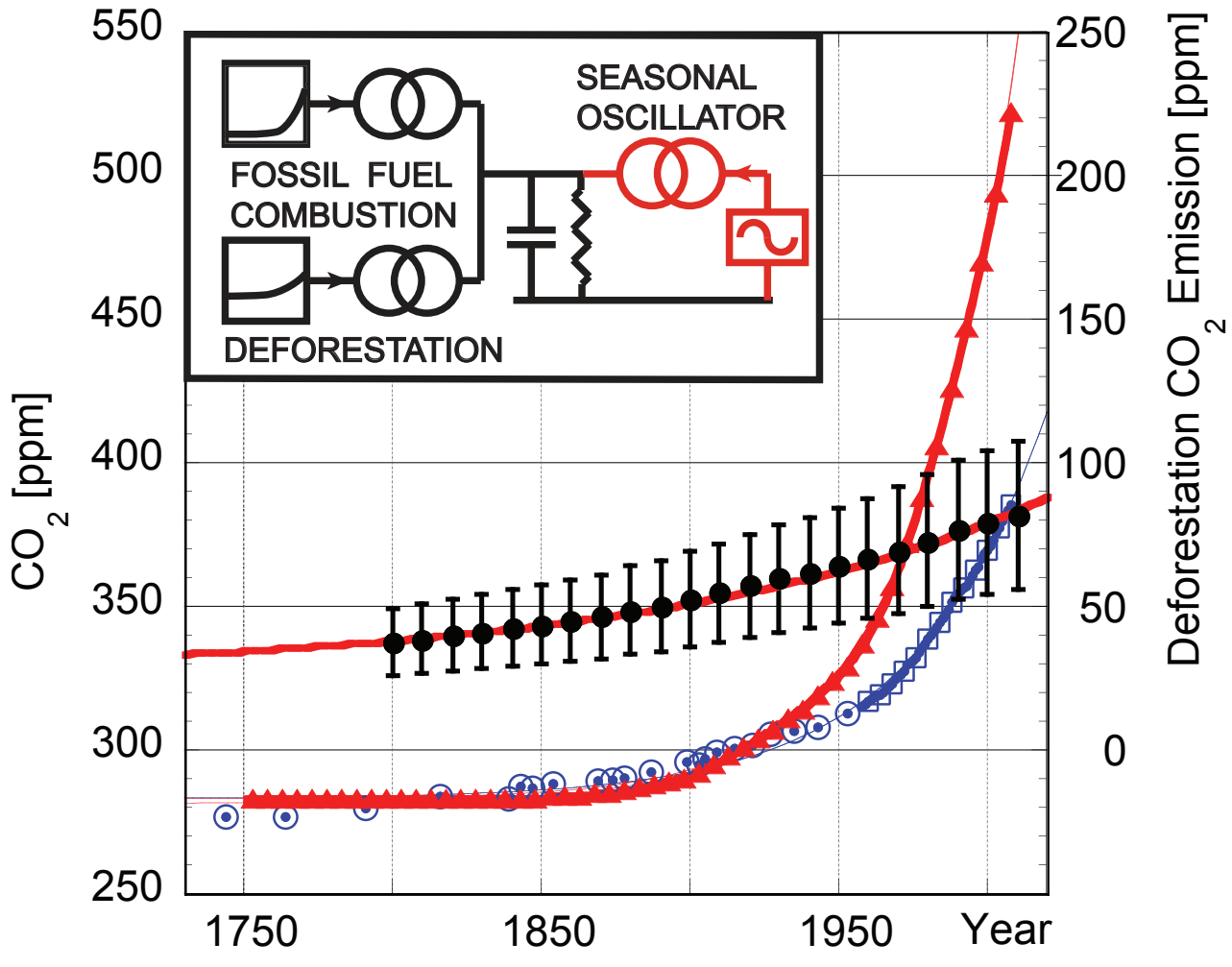

FIG. 1. (Color online) Left scale: $CO_2$ injected by the combustion of fossil fuel (triangles) and annual average of Mauna Loa measurements of atmospheric $CO_2$ concentration (squares). Measurements from ice core samples at the Siple station in Antarctica are added for the pre 1957 period (circles). Right scale: $CO_2$ injected by deforestation (dots with large error bars), note the shifted baseline.

## IV. RETENTION TIME UPPER LIMIT

The atmospheric system can be modeled as the simple electrical integrator circuit shown in the insert of Fig. 1. The time-varying fluxes of $CO_2$ fed by current generators are accumulated in a capacitor that represents the atmosphere, while the resistor accounts for the leakage and retention deficit. Whatever the sink may be, the atmosphere is equivalent to a leaky integrator. The level in an integrator with time constant RC=$\tau_{damp}$ driven by an exponential function with a lifetime $\tau_{drive}$ has an amplitude coefficient controlled by the growth time constant and the retention fraction. The output (retained $CO_2$ concentration) has the same functional shape as the injection, with an amplitude coefficient controlled by the ratio of the two time constants. The evaluation of the atmospheric $CO_2$ lifetime would be straightforward, except for the complication of the slower deforestation contribution, whose amplitude is poorly known. Only an upper limit calculation is possible, starting from the injection time constant of $\tau_{growth}$ = 34.9±0.4 year from the fit in Fig 1. In the last 50 years, the annual ratios of the $CO_2$ concentration and injection increments were r = 0.39±0.02 (which is also an upper limit because deforestation and other contributions are disregarded). An exponential time constant $\tau_{damp} = - \tau_{growth} *$ r /(1-r) = 22.6 year is determined using the leaky integrator model. A 1.2 year error is evaluated from the statistical fluctuations of the retention ratio and the fitting routine uncertainties. The assumption of a sink rate proportional to the concentration, necessary for the validity of the RC circuit model, is validated by the stability of the 0.39 retention ratio over the entire 50 year of useful data, which is further discussed in section VII.

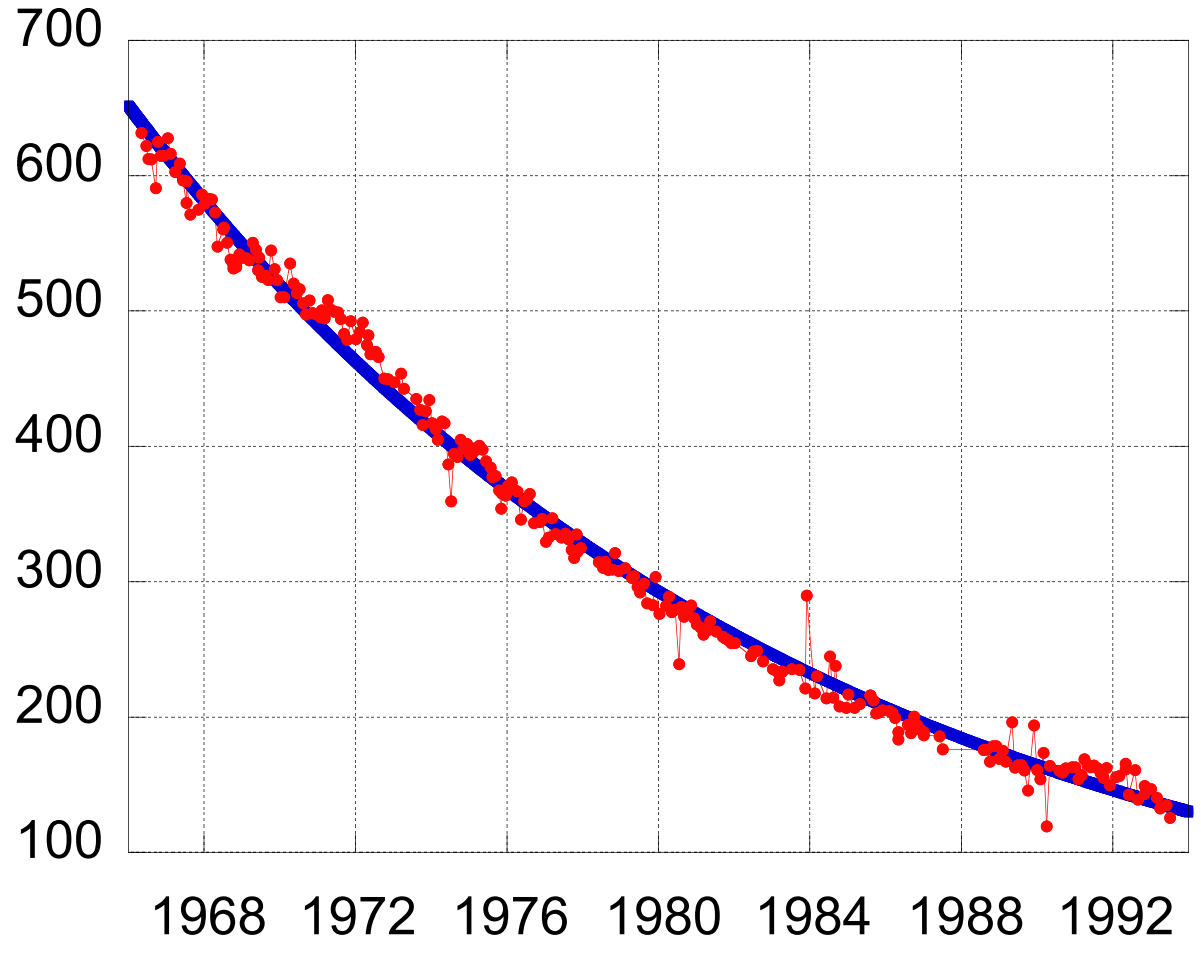

FIG. 2. (Color online) Decay of the concentration of $^{14}C$ after the end of the atmospheric atomic bomb tests.

## V. RETENTION TIME LOWER LIMIT

A lower limit evaluation of atmospheric $CO_2$ lifetime is obtained from a fit to the concentration decay of the $^{14}C$ produced by atmospheric atomic bomb tests (ended in 1967) measured in Wellington (NZ) [4] and shown in Fig 2. One-year delay is observed in the $^{14}C$ peaking time of the Wellington (NZ) with resect to similar measurements performed in Wien (A) [5]. A shortened time constant is expected for a minority gas like $^{14}CO_2$ because of the seasonal fluxes discussed in section VI, and in [2]. Any incomplete return from exchanges of $^{14}C$ with $^{12}C$ molecules in seasonal sinks, for example residues of leaves and other organic matter, accelerates the measurable shedding of atmospheric $^{14}C$. Therefore the exponential decay fit, which gives a decay time of 17.9±0.5 year, can be interpreted as a lower limit of the $CO_2$ retention time constant. The two fits thus bracket the atmospheric $CO_2$ retention lifetime between 22.6 and 17.9 year.

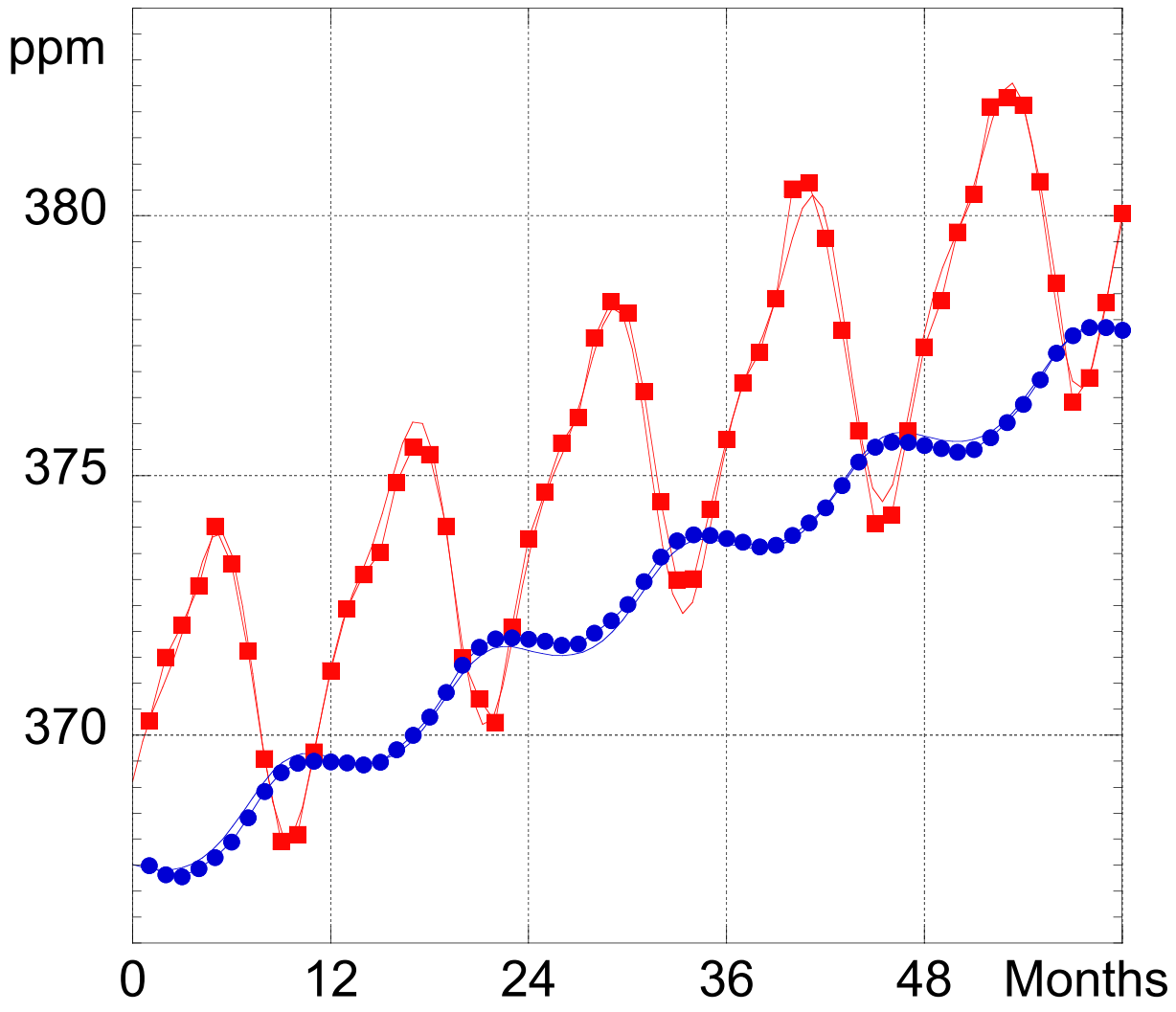

FIG. 3. (Color online) Atmospheric $CO_2$ data between 2001 and 2005 at South Pole (dots) and Mauna Loa (squares). The lines are second harmonic fits added to a linear slope.

## VI. SEASONAL OSCILLATIONS

The seasonal oscillations of atmospheric $CO_2$ concentration first observed by C. Keeling [6] are well visible in the data, they involve large $CO_2$ exchanges with the biota. A subset of the data is shown Fig 3. The oscillations were fit with a 12-month period sine function, and its harmonics to evaluate the oscillation amplitude.

The modulation strength in the Northern hemisphere has a peak flux that exceeds by more than an order of magnitude all human contributions, in agreement with [2]. The modulation is weaker in the Southern hemisphere with a six months phase delay due to the season inversion.

The seasonal variation amplitudes were evaluated for all available stations and plotted as a function of latitude, together with the fraction of Earth covered by landmasses, in Fig 4. Good correlation is evident, with the exclusion of the polar regions, indicating that the periodic sink is mainly terrestrial, in agreement with what stated in [1,section 6.3.2, page 488-489].

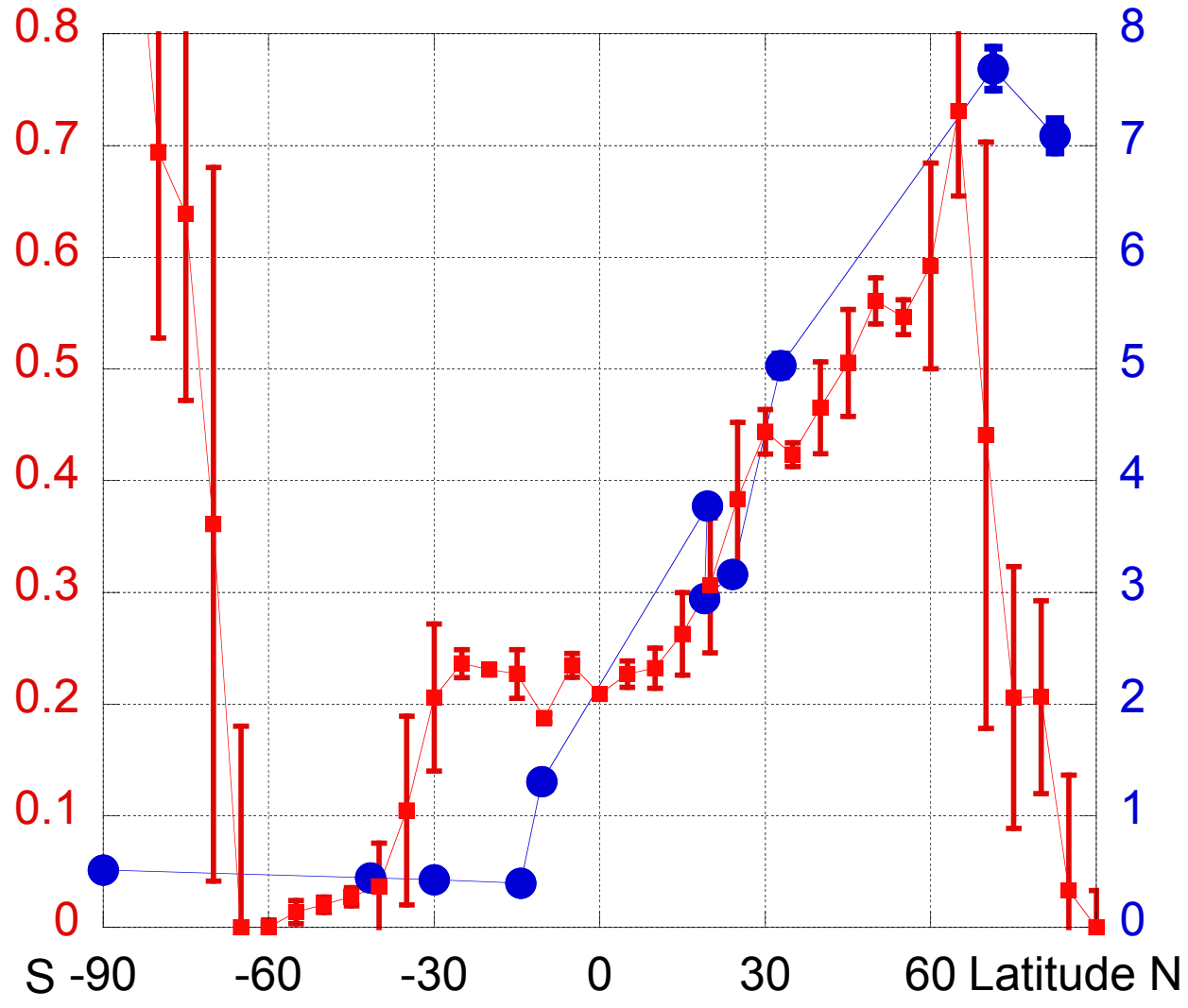

FIG. 4. (Color online) Fraction of Earth covered with landmasses (squares, left scale) versus latitude, and atmospheric $CO_2$ modulation amplitude (dots, right scale). The land coverage error bars are the variation in each of the 5° intervals.

## VII. STABILITY OF SINKS

The monthly Mauna Kea data were used to study the behavior of the modulation amplitude vs. the changing atmospheric $CO_2$ inventory, which is a pre-existing, naturally-occurring exchange cycle. The resulting data is plotted in Fig 5. The 8.8 $10^{-3}$ (±0.5 $10^{-3}$ standard deviation, ±0.1 $10^{-3}$ standard error) amplitude vs. inventory ratio is constant over the years with i.e. the modulation closely tracks the overall atmospheric inventory of $CO_2$. This implies that the modulation remained proportional to the entire $CO_2$ atmospheric inventory, which between 1958 and 2008 grew by ~ 20%.

Similarly, the 60.7% deficit of $CO_2$ (annually evaluated difference between injected and measured concentration) shows no statistically significant change over the last 50 years. The long-term stability of the two ratios imply a, linear mechanism i.e. the absorption (by the biota as a whole and by any other physical process) is proportional to $CO_2$ availability and there is no indication of sink saturation. One could have expected this finding, given the century characteristic times of the long-term sinks listed in [1, table 6.15, page 549]. The stability of sinks is further studied as follows.

A linear fit to the ratio of the modulation amplitude to the atmospheric $CO_2$ inventory presents a statistically negligible but negative slope (-1.7 $10^{-6}$ ± 4.9 $10^{-6}$). A positive slope would be the first sign of saturation of seasonal sinks.

A linear fit to the deficit presents an also statistically negligible positive slope (0.7 $10^{-3}$ ± 1.0 $10^{-3}$). A negative slope would be the first sign of long-term sink saturation. The observed signs, if confirmed by future measurements, would indicate a progressive improvement of the sink's efficiency. The same conclusion could be drawn for the long-term sinks from the exponential fit to atmospheric $CO_2$ that has a longer time constant than the injection rate (fits to data of figure 1).

The statistical analysis of atmospheric $CO_2$ data show no sign of sink saturation.

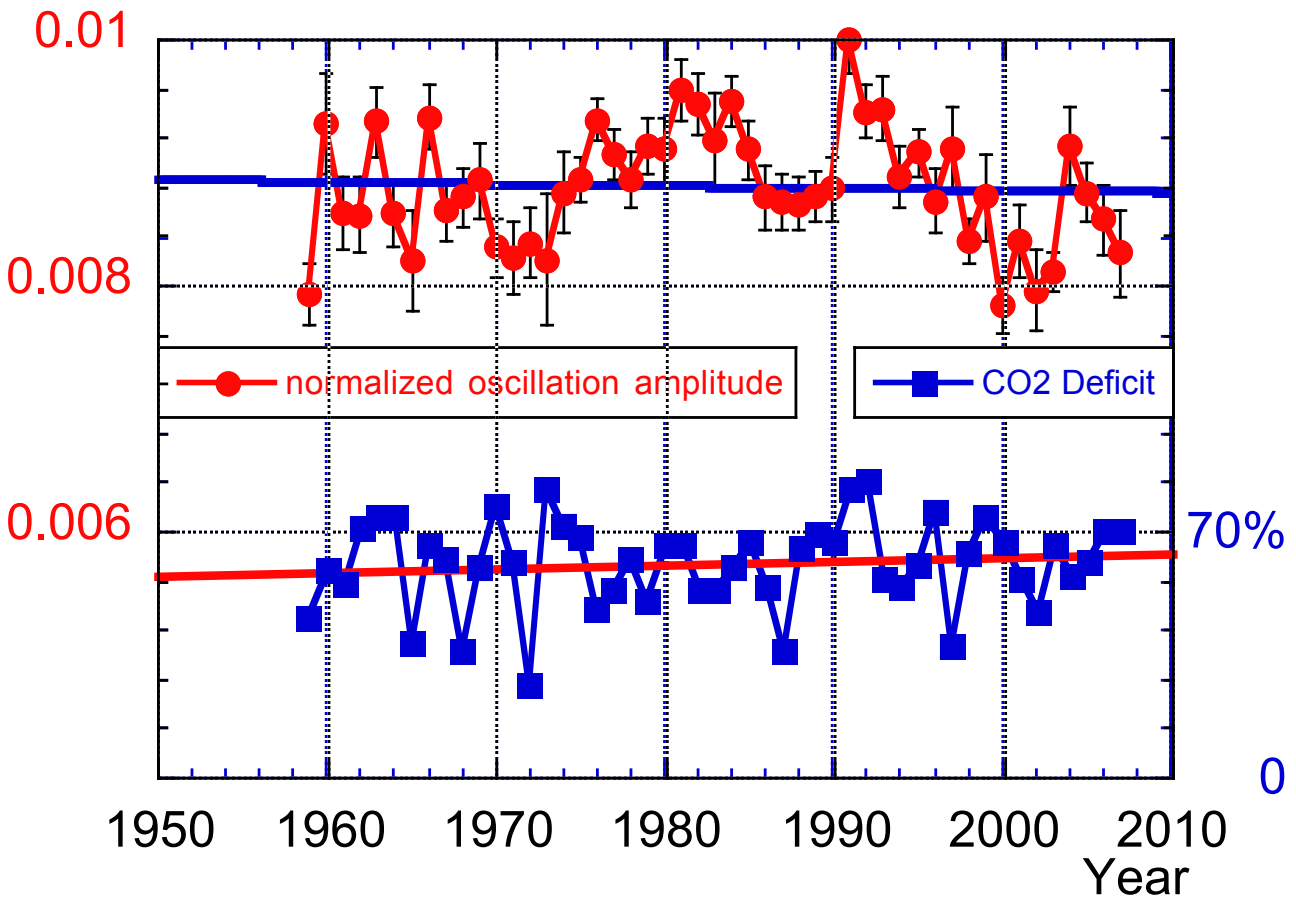

FIG. 5. (Color online) Dots, time dependence of the ratio of the seasonal oscillation amplitude to the atmospheric $CO_2$ inventory, i.e. to the sum of pre-existing and anthropogenic $CO_2$. The error bars represent the fitting errors to each annual oscillation; the fit residuals are <1.7 times the fitting error bars. Squares, fractional deficit of atmospheric $CO_2$. The lines are linear fits to the data.

## VIII. CONCLUSIONS

The 17.9 to 22.6 year time constant here evaluated for the long-term atmospheric $CO_2$ sequestering mechanism implies that any stabilization or reduction of fossil fuel consumption would be reflected in an equally rapid stabilization or reduction of the atmospheric $CO_2$ inventory, and its effects on climate changes. The process is like adding water to a leaky bucket (leaking 61% of each yearly injection); the level rises only if the water flow increases. The growth of atmospheric $CO_2$ is therefore best described as a transit flow through the atmosphere, towards its sequestering sinks. The measured $CO_2$ level grows only because the industrial activities are increasing.

From the atmospheric $CO_2$ observational point of view, during the last half-century the segregation process and the sinks, as a whole, were very stable, which is in agreement with the century-scale characteristic times of the long-term sinks [1, table 6.15, page 549].

It is predicted, in disagreement with [1, box 6.1, page 472], that if the increase of anthropogenic production is halted or reduced the $CO_2$ concentration will stabilize or start reducing within a quarter century.

## ACKNOWLEDGEMENTS

The author would like to thank Patrick Fleury, David Blair, Jan Harms, and Sydney Meshkov for many, very useful discussions and suggestions that greatly improved this paper.